\documentclass{article}
\usepackage{spconf,amsmath,graphicx}
\usepackage{url, times, booktabs, tabularx, amssymb, bm, mathrsfs, mathtools, bbm}
\usepackage{calrsfs, color}
\usepackage[utf8]{inputenc}
\usepackage[final]{microtype}
\usepackage{booktabs}
\usepackage{array}
\usepackage{enumitem}
\usepackage{soul}
\usepackage{cite}
\usepackage{siunitx}
\usepackage{adjustbox}
\usepackage{hhline}
\usepackage{makecell}
\usepackage{caption}
\usepackage{tablefootnote}
\usepackage{scrextend}

\DeclareMathAlphabet{\pazocal}{OMS}{zplm}{m}{n}
\newcolumntype{P}[1]{>{\centering\arraybackslash}p{#1}}
\newcolumntype{?}{!{\vrule width 2pt}}

\title{An Improved Event-Independent Network for \\Polyphonic Sound Event Localization and Detection}
\name{Yin Cao$^{1}$, Turab Iqbal$^{1}$, Qiuqiang Kong$^{2}$, Fengyan An$^{3}$, Wenwu Wang$^{1}$, Mark D. Plumbley$^{1}$}
\address{ $^{1}$Centre for Vision, Speech and Signal Processing (CVSSP), University of Surrey, UK \\
\{yin.cao, t.iqbal, w.wang, m.plumbley\}@surrey.ac.uk \\
$^{2}$ByteDance Shanghai, China, kongqiuqiang@bytedance.com\\
$^{3}$Qingdao University of Technology, China, anfy@qut.edu.cn\\
}

\begin{document}
\ninept
\maketitle
\begin{sloppy}
\begin{abstract}
Polyphonic sound event localization and detection (SELD), which jointly performs sound event detection (SED) and direction-of-arrival (DoA) estimation, detects the type and occurrence time of sound events as well as their corresponding DoA angles simultaneously. We study the SELD task from a multi-task learning perspective. Two open problems are addressed in this paper. Firstly, to detect overlapping sound events of the same type but with different DoAs, we propose to use a trackwise output format and solve the accompanying track permutation problem with permutation-invariant training. Multi-head self-attention is further used to separate tracks. Secondly, a previous finding is that, by using hard parameter-sharing, SELD suffers from a performance loss compared with learning the subtasks separately. This is solved by a soft parameter-sharing scheme. We term the proposed method as Event Independent Network V2 (EINV2), which is an improved version of our previously-proposed method and an end-to-end network for SELD. We show that our proposed EINV2 for joint SED and DoA estimation outperforms previous methods by a large margin, and has comparable performance to state-of-the-art ensemble models.


\end{abstract}
\begin{keywords}
Sound event localization and detection, direction of arrival, event-independent, permutation-invariant training, multi-task learning.
\end{keywords}
\section{Introduction}
\label{sec:intro}

Sound source localization is a challenging research topic \cite{brandstein2013microphone}, with applications in areas such as moving robots, scene visualization systems, and smart homes \cite{virtanen2018computational}. Sound event localization and detection (SELD) estimates the locations of sound sources and detects the corresponding types and occurrence time of the sound events.

For sound event detection (SED), learning-based methods \cite{bishop2006pattern}, which learn models on a dataset, have recently achieved state-of-the-art performance \cite{lin2019guided}. However, for DoA estimation, parametric methods \cite{pavlidi2013real}, which use traditional signal processing algorithms, and learning-based methods, seem to have different strengths. Parametric methods do not need a training dataset, but their generalization ability may not be as good and the number of sources usually needs to be known \textit{a priori} \cite{pavlidi2013real}. Learning-based methods, on the other hand, need a labeled dataset, but can adapt to different complex environments. In this paper, we focus on learning-based methods.

SELD was first introduced in Task 3 of the 2019 Detection and Classification of Acoustic Scenes and Events (DCASE) Challenge, which used the TAU Spatial Sound Events 2019 dataset \cite{Adavanne2019_DCASE, Adavanne2018_JSTSP}. A more challenging dataset with moving sources was released in Task 3 of the 2020 DCASE Challenge \cite{politis2020dataset}. In the datasets for both challenges, there can be up to two overlapping events. For SELD, we previously introduced a two-stage method which detects the sound event first and then transfers the learned representations to extract direction-of-arrival (DoA) features \cite{Cao2019}. While that method achieved a good ranking in DCASE 2019, it intrinsically treats SELD as two separate tasks, without continuously utilizing the essential interactions between SED and DoA estimation. In addition, due to its output format, it is unable to detect different sound events of the same type that have different DoAs. We call this situation \textit{homogeneous overlap}. 

In this paper, we focus on two open issues: the output format and joint SELD learning. We show that, given an effective joint SELD learning scheme, SED and DoA estimation can be trained together with mutual benefits. Source code is released\footnote{\url{https://github.com/yinkalario/EIN-SELD}}.


Firstly, to detect the homogeneous overlap, the trackwise output format is investigated thoroughly. We proposed an Event-Independent Network using a trackwise output format to detect the homogeneous overlap in a previous study \cite{cao2020event}. The trackwise output format assumes that the output of the network has several tracks, each with at most one predicted event with a corresponding DoA. Different tracks can detect events of the same type with different DoAs, which means the trackwise output format can detect the homogeneous overlap \cite{cao2020event}. However, the trackwise output format introduces a \textit{track permutation} problem, which is similar to the talker permutation problem in speaker separation \cite{yu2017permutation, kolbaek2017multitalker}. To address this, we investigate frame-level and chunk-level permutation-invariant training (PIT) to dynamically assign labels to the correct tracks during training. We also propose to use multi-head self-attention (MHSA) from transformers \cite{vaswani2017attention, tay2020efficient} to separate the latent representations. 

%



Secondly, we propose an improved joint-learning method with soft interactions between SED and DoA estimation for the trackwise output format. From a learning perspective, SELD can be considered as a Multi-Task Learning (MTL) problem \cite{ruder2017overview}. MTL is typically done with either hard parameter-sharing (PS) or soft PS. Hard PS means that subtasks use the same high-level feature layers, while soft PS means that different subtasks use feature layers of their own, and there are connections between different feature layers. From previous research \cite{Cao2019, Adavanne2018_JSTSP}, it is found that hard PS used in SELDnet or domain adaptation, which inductively transfers the learned representations from task to task, as used in our previous two-stage method, does not learn an optimal model. Instead, we propose to adopt a soft PS strategy between SED and DoA estimation to learn an optimal model. In particular, by using concepts inspired by the cross-stitch module \cite{misra2016cross}, we incorporate soft PS for both high-level feature layers and the MHSA that is used for track separation.

We propose a method that combines the trackwise output format and the soft PS scheme. We call this method the Event-Independent Network V2 (EINV2). To the best of our knowledge, it is the first time that SELD has been discussed from a MTL perspective. We will show that the proposed method outperforms previous methods by a large margin, and a single EINV2 model using a simple VGG-style architecture gives comparable performance to the state-of-the-art ensemble models on Task 3 of the 2020 DCASE Challenge.


The rest of the paper is arranged as follows. Section \ref{sec:related_works} reviews related works. Section \ref{sec:method} introduces the proposed method. Section \ref{sec:experiments} shows experimental results. Section  \ref{sec:conclusion} summarizes our work.

\section{Related Works}
\label{sec:related_works}

\subsection{Sound Event Localization and Detection}

SELD has received wide attention since Task 3 of 2019 DCASE Challenge \cite{Mesaros2016_MDPI, Adavanne2018_JSTSP, Adavanne2019_DCASE, politis2020overview}. A new TAU-NIGENs Spatial Sound Events dataset with moving sound sources that promotes research in this area was recently released \cite{politis2020dataset, Mesaros_2019_WASPAA}. Adavanne et al. \cite{Adavanne2018_JSTSP} proposed SELDnet, where SED and DoA estimation share high-level feature layers. We proposed a two-stage method by means of domain adaptation \cite{Cao2019}. Grondin et al. \cite{Grondin2019} used a CRNN on pairs of microphones to perform SELD. Nguyen et al. \cite{nguyen2020sequence} proposed to use a sequence matching network to align SED and DoA predictions. Mazzon et al. \cite{Mazzon2019} proposed a spatial-augmentation method by rotating channels. Shimada et al. \cite{shimada2020sound} proposed an Activity-Coupled Cartesian DoA Vector (ACCDOA) method to train only on location information but using magnitudes of ACCDOA vectors as SED activations. In this paper, an improved Event-Independent Network V2 using trackwise output format and a soft PS scheme is proposed.


\subsection{Multi-Task Learning}

MTL has been successfully applied to almost all areas of machine learning \cite{ruder2017overview, zhang2017survey , zhang2018learning}, such as natural language processing \cite{subramanian2018learning} and computer vision \cite{zamir2018taskonomy}. MTL is inherently a multi-objective problem with conflicts existing among tasks \cite{sener2018multi}. A weighted linear combination of per-task losses is a common balanced solution. However, this simple solution may be invalid and make MTL detrimental when tasks heavily compete. Therefore, the way that features are shared among tasks is an essential problem. Some methods have been investigated, e.g. fully-adaptive feature sharing\cite{lu2017fully}, joint many-task model \cite{hashimoto2017joint}, Panoptic Feature Pyramid Networks \cite{kirillov2019panoptic}, and so on. Previous experimental results show that SED and DoA share some common features, but also compete, which makes the hard PS used by SELDnet \cite{Adavanne2018_JSTSP} a suboptimal solution. In this paper, a soft PS method is used.


\section{The Proposed Method}
\label{sec:method}


\subsection{SELDnet Output Format}

We first review the SELDnet output format \cite{Adavanne2018_JSTSP}. An illustration of the output format is given in Fig. \ref{fig:output_formats}. Mathematically the SELDnet output format is defined as


\vspace*{-4mm}
\begin{equation}
\label{eqn:seldnet_format}
\boldsymbol{Y}_{\text{SELDnet}}=\{ \left({y}_\mathrm{SED}, {y}_\mathrm{DoA}\right) | {y}_\mathrm{SED} \in \mathbbm{1}_\mathbf{S}^{K}, {y}_\mathrm{DoA} \in \mathbb{R}^{K \times 3} \},
\end{equation}
\vspace*{-4mm}

\noindent where ${y}_\mathrm{SED}$ and ${y}_\mathrm{DoA}$ are predictions for SED and DoA, respectively, $\mathbbm{1}_\mathbf{S}^{K}$ is the one hot encoding for $K$ classes, $\mathbf{S}$ is the set of sound event classes, and the number of dimensions of Cartesian coordinates is $3$.

\begin{figure}[tb]
  \centering
  \scalebox{1.0}{\centerline{\includegraphics[width=\columnwidth]{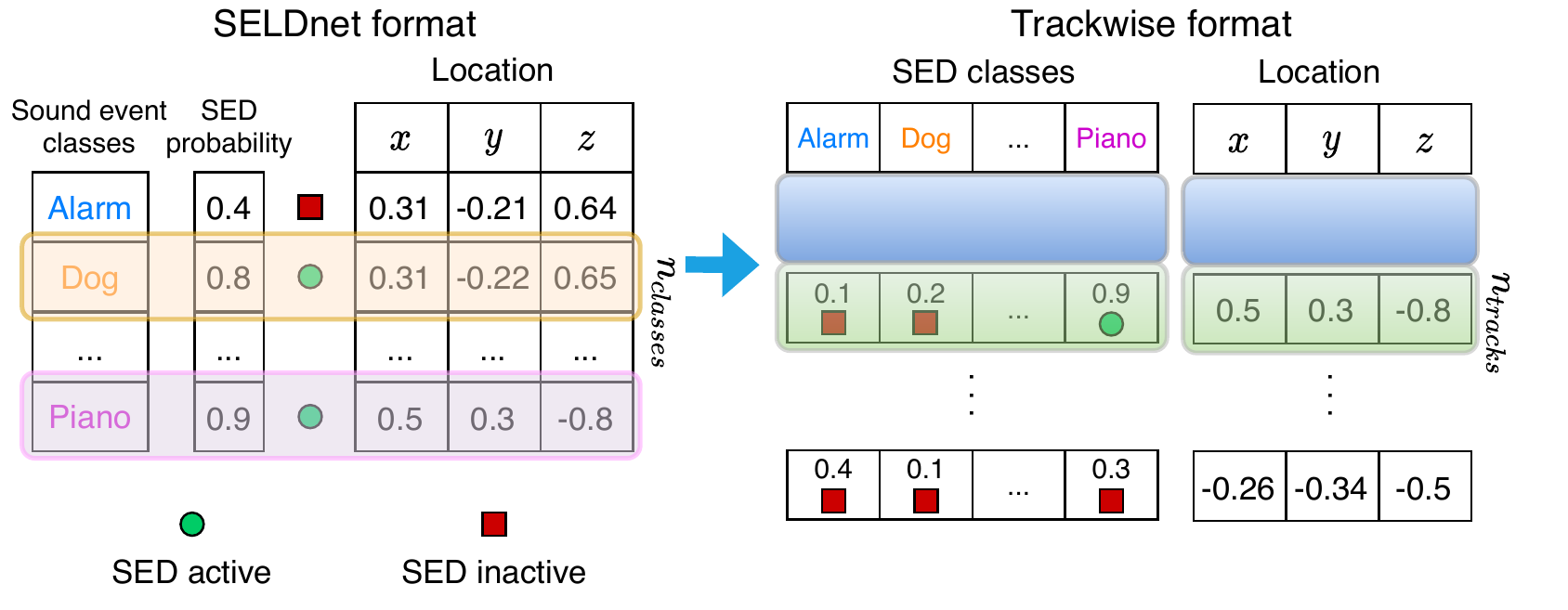}}}
  \vspace*{-5mm}
  \caption{Illustration of output formats, from SELDnet to Trackwise.}
  \label{fig:output_formats}
\vspace*{-4mm}
\end{figure}


SELDnet predicts probabilities of all sound events and corresponding locations. A threshold is then used to binarize the probabilities of events. Binarized probabilities are used to activate corresponding sound events and locations. The number of predicted event types and locations are the same (their dimensions are both $K$), which means that there is only one location per predicted event. 

Due to the fact that only one location can be predicted for one event, the SELDnet output format cannot detect the same event with multiple locations, hence it is invalid when there is a homogeneous overlap. In addition, we found that when only a few types of sound events are active, locations of other inactive events tend to be very similar to locations of activated events. For instance, in Fig. \ref{fig:output_formats}, the SELDnet output format shows that the event ``Alarm'' is not active, but its location is very similar to the location of ``Dog''. Considering the fact that there is only a limited number of active sources at frame $t$, predicting locations for all of the event types, no matter if they are active or not, is unnecessary. Therefore, dimensions of location are redundant. Regression naturally fits when estimating continuous locations. Redundant dimensions of regression may increase the demand for the training-data size and the model capacity. 

\subsection{Trackwise Output Format}
\label{ssec:track_wise_format}

The trackwise output format was first proposed in our previous work \cite{cao2020event} and is also shown in Fig. \ref{fig:output_formats}. It can be defined as

\vspace*{-4mm}
\begin{equation}
\label{eqn:trackwise_format}
\resizebox{0.9\hsize}{!}{
$\boldsymbol{Y}_{\text{Trackwise}}=\{ \left({y}_\mathrm{SED}, {y}_\mathrm{DoA}\right) | {y}_\mathrm{SED} \in \mathbbm{1}_\mathbf{S}^{M \times K}, {y}_\mathrm{DoA} \in \mathbb{R}^{M \times 3} \}$
},
\end{equation}
\vspace*{-4mm}

\noindent where $M$ is the number of tracks. Since $K$ is the total sound events categories, $M \ll K$ in general.

Each track only detects one event and a corresponding location. The number of tracks $M$ is the desired number of DoAs to detect. Hence, instead of estimating locations for all of the $K$ events regardless of whether they are active or not, the trackwise output format only estimates $M$ locations. This greatly reduces the demand for the model capacity and the required data size. In addition, the output format can now detect the homogeneous overlap.


The trackwise output format introduces a \textit{track permutation} problem \cite{cao2020event}. Since $M \ll K$, events are not always predicted in fixed tracks. At frame $t$, ``Dog'' can be predicted in track 1, but at next frame $t+1$, ``Dog'' can be predicted in track 2. The prediction of ``Dog'' is not fixed on track 1 or 2. This would result in a consequence that tracks do not ``know'' the correct ground truth during training. Permutation-invariant training is used to solve this problem.


\subsection{Permutation-Invariant Training}
\label{ssec:pit}

Permutation-invariant training (PIT) was first proposed for speaker separation \cite{yu2017permutation}. Our previous work discussed the benefit of using PIT for the trackwise output format \cite{cao2020event}. Frame-level and chunk-level PIT are both used in this paper. Here, chunk is a whole segment spanning from the start to the end of an event. Frame-level PIT assumes labels among frames are independently assigned, chunk-level PIT assumes labels that are within the same chunk of audio event are assigned to the same track. Let $o$ denote the frame index $t$ or the chunk index $c$. Given a permutation set $\mathbf{P}(o)$ consisting of all possible prediction-label pairs at index $o$, ground truth labels are assigned using all possible combinations in $\mathbf{P}(o)$. The lowest loss for each frame (tPIT) or for each chunk (cPIT) will be chosen to perform the back-propagation. The PIT loss can be defined as

\begin{equation}
\label{eqn:pit}
    \pazocal{L}^{PIT}(o)=\min _{\alpha \in \mathbf{P}(o)} \sum_{M} \{ \ell^{\text{SED}}_{\alpha}(o) + \ell^{\text{DoA}}_{\alpha}(o) \},
\end{equation}
\vspace*{-2mm}

\noindent where $\alpha \in \mathbf{P}(o)$ is one of the possible permutation pairs. 


\subsection{Multi-Head Self-Attention}
\label{ssec:mhsa}

The MHSA in transformers \cite{vaswani2017attention} is used to separate tracks. A fixed absolute positional encoding is used before MHSA as:

\vspace*{-4mm}
\begin{equation}
\resizebox{0.9\hsize}{!}{
$\boldsymbol{P}_{(t, 2 i)} = 0.1 \sin \left(t / 10^{8 i / D_{c}}\right), \;\; \boldsymbol{P}_{(t, 2 i+1)} = 0.1 \cos \left(t / 10^{8 i / D_{c}}\right)$,
}
\end{equation}
\vspace*{-4mm}

\noindent where $t$ denotes the index of the time dimension, and $i$ denotes the index of the feature maps. Given an input $\boldsymbol{X} \in \mathbb{R}^{D_{t} \times D_\mathrm{in}}$ with $D_\mathrm{in}$ denoting the input dimension, the Self-Attention (SA) can be written as:

\vspace*{-4mm}
\begin{equation}
\resizebox{0.9\hsize}{!}{
$\text { SA }(\boldsymbol{X}):=\operatorname{softmax}\left( (\boldsymbol{X}+\boldsymbol{P}) \boldsymbol{W}_\mathrm{q r y} \boldsymbol{W}_\mathrm{k e y}^{\top}(\boldsymbol{X}+\boldsymbol{P})^{\top} \right) \boldsymbol{X} \boldsymbol{W}_\mathrm{val}$,
}
\end{equation}
\vspace*{-4mm}

\noindent where $\boldsymbol{W}_\mathrm{q r y}, \boldsymbol{W}_\mathrm{k e y} \in \mathbb{R}^{D_\mathrm{i n} \times D_{k}}$ are learnable query and key matrices, respectively, $D_{k}$ is the dimension of keys, $\boldsymbol{W}_\mathrm{val} \in \mathbb{R}^{D_\mathrm{i n} \times D_\mathrm{out}}$ is a learnable value matrix. MHSA evenly splits $D_\mathrm{out}$ to $N_{h}$ head, with each head having a dimension of $D_{h}$. MHSA can be expressed as:

\vspace*{-2mm}
\begin{equation}
\resizebox{0.7\hsize}{!}{
$ \operatorname{MHSA}(\boldsymbol{X}):=\underset{h \in\left[N_{h}\right]}{\operatorname{concat}}\left[\text { SA }_{h}(\boldsymbol{X})\right] \boldsymbol{W}_\mathrm{out}+\boldsymbol{b}_\mathrm{out} $,
}
\end{equation}

\noindent where $\boldsymbol{W}_\mathrm{out} \in \mathbb{R}^{N_{h} \cdot D_{h} \times D_\mathrm{out}}$ and $\boldsymbol{b}_\mathrm{out} \in \mathbb{R}^{D_\mathrm{out}}$ are the projection matrix and the corresponding bias, respectively. 



\subsection{Parameter-Sharing Strategies}
\label{ssec:joint_seld}

From an MTL perspective, joint SELD learning can be mutually beneficial \cite{ruder2017overview, zhang2017survey}. The explanation is three-fold: (i) SED and DoA estimation have different noise patterns. When training a model on both SED and DoA estimation, the idea is to learn a good representation $F$ that generalizes well through averaging both data-dependent or label-dependent noise patterns; (ii) some features $R$ may be easy to learn for SED, while being hard to learn for DoA estimation. Using MTL, the model can eavesdrop to learn features $R$ through SED; (iii) MTL uses several loss terms, which may act in the same way as loss regularizers. Hence, MTL can also reduce the overrfit of the model to SED or DoA estimation. Hard and soft PS are two typical methods to implement MTL. Hard PS means subtasks use the same feature layers, whereas soft PS means subtasks use their own feature layers with connections existing among those feature layers.


As shown in Fig. \ref{fig:parameter_sharing}, three PS strategies are given for the trackwise output format. Fig. \ref{fig:parameter_sharing}(a) shows hard PS for the trackwise output format. High-level representations are fully shared between two tasks. However, it is inevitable that as two different tasks, SED and DoA estimation conflict. Hard PS forces them to use the same features, which leads to a performance loss compared with learning the subtasks separately. Fig. \ref{fig:parameter_sharing}(b) shows no PS with the trackwise output format. This method treats two tasks as separate ones. Since SED and DoA labels are connected, there are still weak connections between the two tasks when they are trained together. Fig. \ref{fig:parameter_sharing}(c) shows soft PS using cross-stitch with the trackwise output format. The cross-stitch units are applied between feature layers and MHSA layers in both SED and DoA estimation. Let $D_{c}, D_{t}, D_{f}$ denote the dimensions of feature maps, time, and frequency, respectively. ${\alpha}_{ij} \in \mathbb{R}^{D_{c}}$ denotes learnable parameters. The new feature maps $ (\bm{\hat{x}}^\mathrm{SED}, \bm{\hat{x}}^\mathrm{DoA}) \in \mathbb{R}^{D_{c} \times D_{t} \times D_{f}}$ can be calculated from the original feature maps $ (\bm{x}^\mathrm{SED}, \bm{x}^\mathrm{DoA}) $ as:


\vspace*{-2mm}
\begin{equation}
\label{eq:soft-connect}
\resizebox{0.65\hsize}{!}{
    $\left[ \bm{\hat{x}}^\mathrm{SED}, \bm{\hat{x}}^\mathrm{DoA} \right] ^ {\top} =  \bm{\alpha} \left[ \bm{x}^\mathrm{SED}, \bm{x}^\mathrm{DoA} \right] ^ {\top}$,
}
\end{equation}
\vspace*{-4mm}

\noindent where $\bm{\alpha}$ is a $2 \times 2$ matrix that consists of elements ${\alpha}_{ij}$, and $[\cdot]^{\top}$ denotes the transpose operation.

\begin{figure}[tb]
    \begin{minipage}[b]{.48\linewidth}
      \centering
      \centerline{\includegraphics[width=4.3cm]{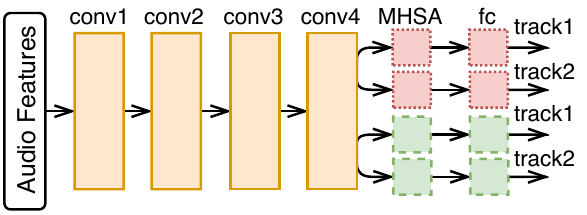}}
      \centerline{(a) Hard parameter-sharing}\medskip
    \end{minipage}
    \hfill
    \begin{minipage}[b]{0.48\linewidth}
      \centering
      \centerline{\includegraphics[width=4.3cm]{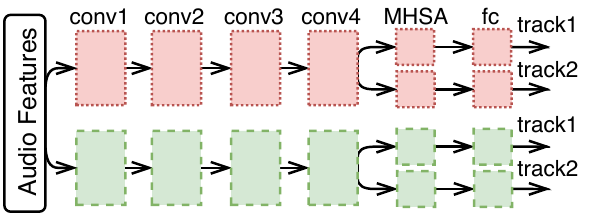}}
      \centerline{(b) No parameter-sharing}\medskip
    \end{minipage}
    \begin{minipage}[b]{1.0\linewidth}
      \centering
      \centerline{\includegraphics[width=8.5cm]{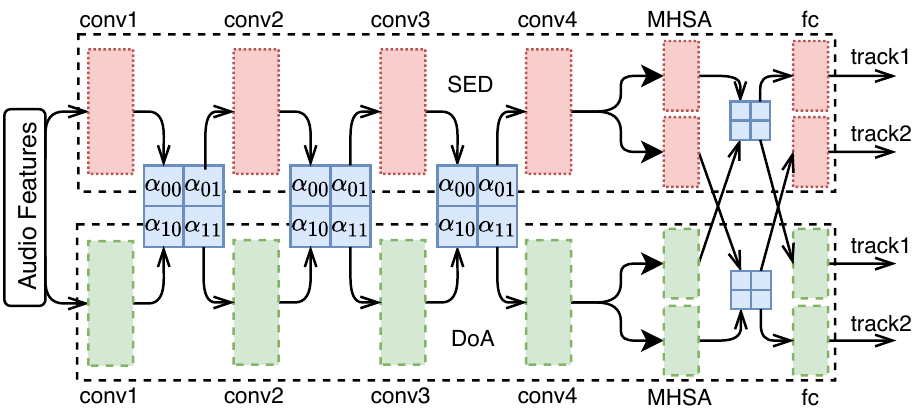}}
      \centerline{(c) Soft parameter-sharing using cross-stitch}
    \end{minipage}
\caption{Different strategies of PS. Dotted-red is the SED task. Dashed-green is the DoA estimation task. Rectangular-blue boxes indicate soft connections between SED and DoA estimation.}
\label{fig:parameter_sharing}
\vspace*{-2mm}
\end{figure}

\begin{table}[tb]
\centering
\caption{Event Independent Network V2}
\label{table:einv2}
\vspace*{-2mm}
\begin{adjustbox}{width=\columnwidth,center}
\begin{tabular}{cccc}
    \Xhline{2\arrayrulewidth}
    \multicolumn{2}{c|}{SED task, log-mel spectrogram} & \multicolumn{2}{c}{DoA estimation task, log-mel + intensity vector}\\
    \hline
    \multicolumn{2}{c|}{$ (3 \times 3 \ @ \ 64 \text{, BN, ReLU}) \times 2 $, Pooling $ 2 \times 2 $} & \multicolumn{2}{c}{$ (3 \times 3 \ @ \ 64  \text{, BN, ReLU}) \times 2 $, Pooling $ 2 \times 2 $} \\ 
    \hline
    \multicolumn{4}{c}{Soft Parameter-Sharing, $ \bm{\alpha}_{2 \times 2} \ @ \ 64 $} \\ 
    \hline
    \multicolumn{2}{c|}{$ (3 \times 3 \ @ \ 128 \text{, BN, ReLU}) \times 2 $, Pooling $ 2 \times 2 $} & \multicolumn{2}{c}{$ (3 \times 3 \ @ \ 128 \text{, BN, ReLU}) \times 2 $, Pooling $ 2 \times 2 $} \\ 
    \hline
    \multicolumn{4}{c}{Soft Parameter-Sharing, $ \bm{\alpha}_{2 \times 2} \ @ \ 128 $} \\ 
    \hline
    \multicolumn{2}{c|}{$ (3 \times 3 \ @ \ 256 \text{, BN, ReLU}) \times 2 $, Pooling $ 1 \times 2 $} & \multicolumn{2}{c}{$ (3 \times 3 \ @ \ 256 \text{, BN, ReLU}) \times 2 $, Pooling $ 1 \times 2 $} \\ 
    \hline
    \multicolumn{4}{c}{Soft Parameter-Sharing, $ \bm{\alpha}_{2 \times 2} \ @ \ 256 $} \\ 
    \hline
    \multicolumn{2}{c|}{$ (3 \times 3 \ @ \ 512 \text{, BN, ReLU}) \times 2 $, Pooling $ 1 \times 2 $} & \multicolumn{2}{c}{$ (3 \times 3 \ @ \ 512 \text{, BN, ReLU}) \times 2 $, Pooling $ 1 \times 2 $} \\ 
    \hline
    \multicolumn{2}{c|}{Global average pooling @ frequency} & \multicolumn{2}{c}{Global average pooling @ frequency} \\
    \hhline{==|==}
    \multicolumn{1}{c|}{Track 1, SED} & \multicolumn{1}{c|}{Track 1, DoA} & \multicolumn{1}{c|}{Track 2, SED} & \multicolumn{1}{c}{Track 2, DoA} \\ 
    \hline
    \multicolumn{1}{c|}{$ ( \text{MHSA} \ @ \ 512 , 8 \ \text{h} ) \times 2 $} & \multicolumn{1}{c|}{$ ( \text{MHSA} \ @ \ 512 , 8 \ \text{h} ) \times 2 $} & \multicolumn{1}{c|}{$ ( \text{MHSA} \ @ \ 512 , 8 \ \text{h} ) \times 2 $} & \multicolumn{1}{c}{$ ( \text{MHSA} \ @ \ 512 , 8 \ \text{h} ) \times 2 $}\\ 
    \hline
    \multicolumn{2}{c|}{Soft Parameter-Sharing, $ \bm{\alpha}_{2 \times 2} \ @ \ 512 $} & \multicolumn{2}{c}{Soft Parameter-Sharing, $ \bm{\alpha}_{2 \times 2} \ @ \ 512 $} \\ 
    \hline
    \multicolumn{1}{c|}{FC, $512 \times 14$, Sigmoid} & \multicolumn{1}{c|}{FC, $512 \times 3$, Tanh} & \multicolumn{1}{c|}{FC, $512 \times 14$, Sigmoid} & \multicolumn{1}{c}{FC, $512 \times 3$, Tanh}\\ 
    \hline
    \multicolumn{1}{c|}{Binary Cross-Entropy} & \multicolumn{1}{c|}{Mean Square Error} & \multicolumn{1}{c|}{Binary Cross-Entropy} & \multicolumn{1}{c}{Mean Square Error}\\ 
    \hline
    \multicolumn{4}{c}{Frame-Level or Chunk-Level permutation-invariant Training}\\ 
    \Xhline{2\arrayrulewidth}
\end{tabular}
\end{adjustbox}
\vspace*{-6mm}
\end{table}

\subsection{Event-Independent Network V2}
\label{ssec:einv2}

The proposed EINV2 combines the trackwise output format, PIT, MHSA, and soft PS. Table \ref{table:einv2} shows the architecture of EINV2. The audio features used are log-mel spectrograms and intensity vectors in mel-space \cite{cao2020event}. The trackwise output format can detect DoAs using only necessary dimensions. It can also detect the homogeneous overlap. PIT is used to solve the track permutation problem. MHSA is used to separate tracks. The soft PS using cross-stitch units is created between latent representations of SED and DoA estimation subtasks, the network can decide what useful information to exchange and what not to. This avoids a performance loss for both subtasks. 



\section{Experiments}
\label{sec:experiments}

The dataset used is TAU-NIGENS Spatial Sound Events 2020 \cite{politis2020dataset}, which consists of 14 types of sound events with continuous DoA angles spanning from $\phi \in[-180,180)$ in azimuth and $\theta \in[-45,45]$ in elevation. There are up to two overlapping events, hence $M$ is set to 2. Four evaluation metrics are used \cite{Mesaros_2019_WASPAA}, F-score $F_{\leq T^{\circ}}$, Error Rate $E R_{\leq T^{\circ}}$, localization error $L E_{\mathrm{CD}}$ and localization recall $L R_{\mathrm{CD}}$. $F_{\leq T^{\circ}}$ and $E R_{\leq T^{\circ}}$ consider true positives predicted under a distance threshold $T=20^{\circ}$ from the ground truth.

\subsection{Hyper-Parameters}

A \num{1024}-point Hann window with a hop size of \num{600} points is used for FFT. The number of mel bands is \num{256}. Audio clips are segmented to have a fixed length of 4 seconds without overlapping for both training and test sets. The AdamW optimizer is used. The learning rate is set to \num{0.0005} for the first 90 epochs and is adjusted to 0.00005 for next 10 epochs that follows. The threshold for SED is \num{0.5} to binarize predictions. All experimental scores are trained on the development set and tested on the evaluation set based on 1 second segments. Final scores are averaged on five different trials.

\begin{figure}[t]
    \begin{minipage}[b]{.42\linewidth}
        \centering
        \captionsetup{type=table}
        \caption{Format comparison.}
        \label{tab:format_comp}
        \begin{adjustbox}{width=\columnwidth,center}
        \begin{tabular}{P{2cm}P{2cm}P{2cm}}
                \Xhline{3\arrayrulewidth}
                \multicolumn{1}{c}{Method}                      & \multicolumn{1}{c}{\begin{tabular}{@{}c@{}}$SED_\mathrm{only}$ \\ $F$\end{tabular}} & \multicolumn{1}{c}{\begin{tabular}{@{}c@{}}$DoA_\mathrm{only}$ \\ $LE$\end{tabular}} \\
                \midrule
                \multicolumn{1}{c}{SELDnet}                     & \multicolumn{1}{c}{\textbf{0.839}}        & \multicolumn{1}{c}{17.21} \\
                \multicolumn{1}{c}{$\text{Track}_\mathrm{noPIT}$} & \multicolumn{1}{c}{0.795}        & \multicolumn{1}{c}{37.57} \\
                \multicolumn{1}{c}{$\text{Track}_\mathrm{cPIT}$}  & \multicolumn{1}{c}{0.832}        & \multicolumn{1}{c}{12.55} \\
                \multicolumn{1}{c}{$\text{Track}_\mathrm{tPIT}$}  & \multicolumn{1}{c}{\textbf{0.838}}        & \multicolumn{1}{c}{\textbf{12.39}} \\
                \Xhline{3\arrayrulewidth}
        \end{tabular}
        \end{adjustbox}
    \end{minipage}
    \hfill
    \begin{minipage}[b]{.56\linewidth}
        \centering
        \centerline{\includegraphics[width=5cm]{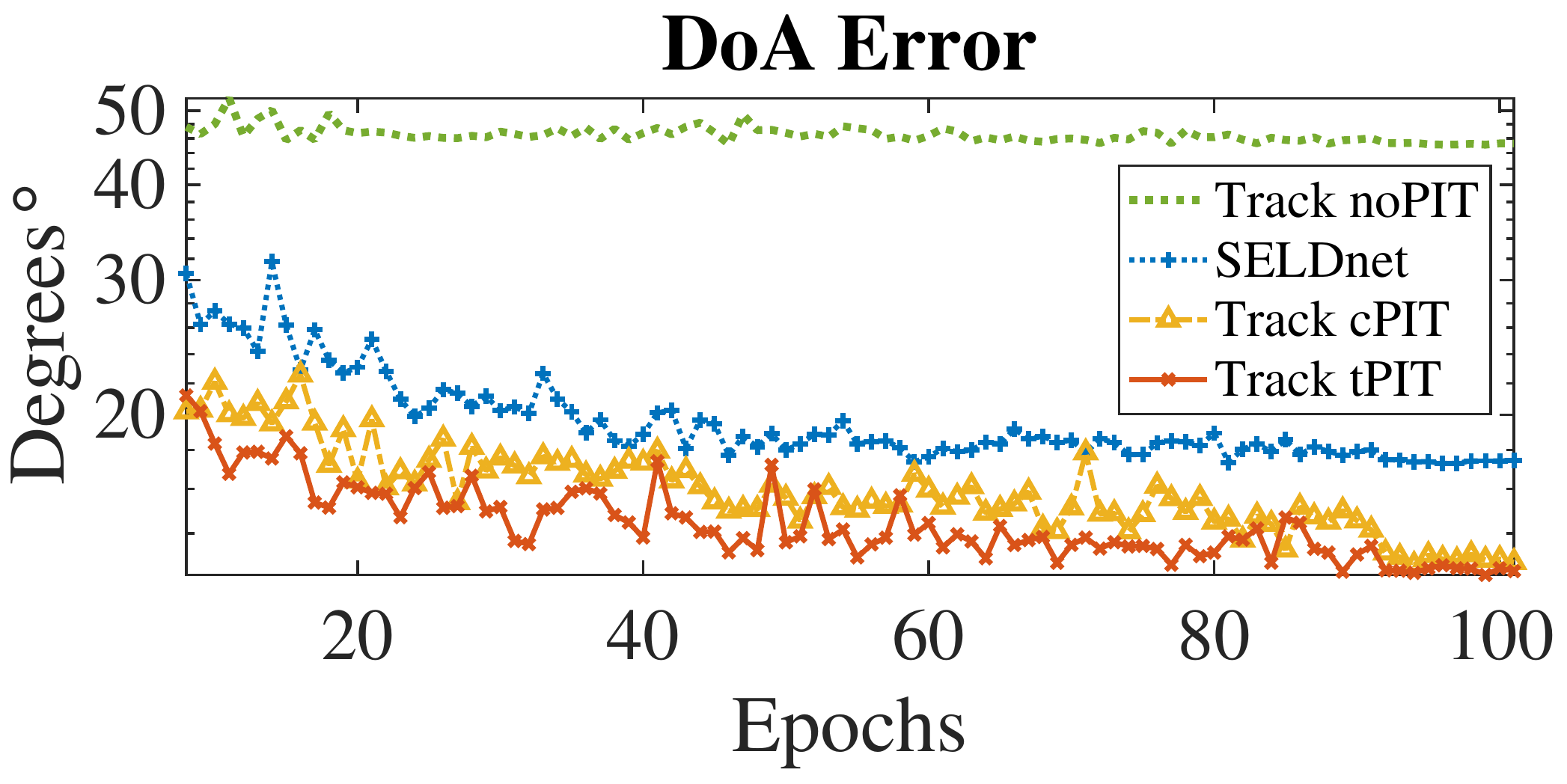}}
        \vspace*{-3mm}
        \caption{Format performance.}\label{fig:format_perf}
        \vspace*{-12mm}
    \end{minipage}
    \vspace*{2mm}
\end{figure}

\begin{table}[t]
    \centering
    \caption{Performance on combined SELD task.}
    \label{tab:perf_seld}
    \vspace*{-2mm}
    \begin{adjustbox}{width=\columnwidth,center}
    \begin{tabular}{cccccc}
            \Xhline{3\arrayrulewidth}
            \multicolumn{1}{c}{Methods}  & \multicolumn{1}{c}{Output Format} & \multicolumn{1}{c}{$E R_{\leq 20^{\circ}}$} & \multicolumn{1}{c}{$F_{\leq 20^{\circ}}$} & \multicolumn{1}{c}{$L E_{\mathrm{CD}}$} & \multicolumn{1}{c}{$L R_{\mathrm{CD}}$} \\
            \midrule
            Baseline                        & SELDnet   & 0.720 & 37.4\% & 22.8$^{\circ}$ & 60.7\% \\
            \midrule
            No-PS                           & SELDnet   & 0.399 & 67.5\% & 14.8$^{\circ}$ & 83.8\% \\
                                            & Trackwise & 0.340 & 73.7\% & 11.9$^{\circ}$ & 83.8\% \\           
            \midrule
            Hard-PS                         & SELDnet   & 0.414 & 65.9\% & 15.5$^{\circ}$ & 82.4\% \\
                                            & Trackwise & 0.339 & 73.9\% & 10.4$^{\circ}$ & 81.3\% \\
            \midrule
            ACCDOA\cite{shimada2020sound}   & SELDnet   & 0.333 & 75.9\% & 10.3$^{\circ}$ & 82.1\% \\
            \midrule
            Soft-PS                         & SELDnet   & 0.350 & 72.1\% & 13.1$^{\circ}$ & 83.8\%\\
            \textbf{EINV2}                  & Trackwise & \textbf{0.299 $\pm$ 0.03} & \textbf{77.0 $\pm$ 0.3}\% & \textbf{8.9 $\pm$ 0.2}$^{\circ}$ & \textbf{83.8 $\pm$ 0.2}\% \\
            \Xhline{3\arrayrulewidth}
            No-PS-DA                        & Trackwise & 0.323 & 75.8\% & 10.2$^{\circ}$ & 83.1\% \\
            Hard-PS-DA                      & Trackwise & 0.264 & 80.5\% & 7.9$^{\circ}$  & 84.3\% \\
            ACCDOA-DA                       & SELDnet   & 0.293 & 80.0\% & 8.7$^{\circ}$  & 84.4\% \\
            \textbf{EINV2-DA}               & Trackwise & \textbf{0.233 $\pm$ 0.03} & \textbf{83.2 $\pm$ 0.2}\% & \textbf{6.8 $\pm$ 0.2}$^{\circ}$ & \textbf{86.1 $\pm$ 0.3}\% \\
            \textit{ACCDOA Ensemble Model}\tablefootnote{\label{note2}\url{https://bit.ly/31edoqC}} & \textit{SELDnet} & $\textit{0.25}$ & $\textit{83.2}\%$ & $\textit{7.0}^{\circ}$ & $\textit{86.2}\%$ \\
            \textit{Best Ensemble Model}\footref{note2} & $-$ & $\textit{0.20}$ & $\textit{84.9}\%$ & $\textit{6.0}^{\circ}$ & $\textit{88.5}\%$ \\
            \Xhline{3\arrayrulewidth}
    \end{tabular}
    \end{adjustbox}
    \vspace*{-6mm}
\end{table}

\subsection{Comparison of the trackwise and the SELDnet formats}
The trackwise and the SELDnet formats are discussed first. To accurately compare the two formats, the single SED and DoA estimation tasks are tested individually. When testing SED alone, DoA predictions are set to be ground truth. A similar setup applies for DoA estimation. Both tasks use a single branch of EINV2 without soft PS shown in Table \ref{table:einv2}. Comparison results and the convergence behaviour of location error are shown in Table \ref{tab:format_comp} and Fig. \ref{fig:format_perf}, respectively. $\text{Track}_\mathrm{noPIT}$, $\text{Track}_\mathrm{cPIT}$ and $\text{Track}_\mathrm{tPIT}$ represents the trackwise output format without PIT, with chunk-level PIT, and with frame-level PIT, respectively. It can be seen that $\text{Track}_\mathrm{tPIT}$ achieves the lowest location error with $\text{Track}_\mathrm{cPIT}$ falling behind by $0.16^{\circ}$. SELDnet gets approximately $5^{\circ}$ higher location error. $\text{Track}_\mathrm{noPIT}$ does not converge. It is within expectations that the SELDnet output format has a higher location error than $\text{Track}_\mathrm{tPIT}$, which reduces redundant dimensions for regression as discussed in Section \ref{ssec:track_wise_format}. When the trackwise output format is used, the track permutation problem prevents the correct labels from being assigned to the corresponding tracks. This can be elegantly solved using PIT. cPIT is more effective at tracking chunk-level events. While the metrics used in this paper are frame-level, which may be the reason why $\text{Track}_\mathrm{tPIT}$ is slightly better than $\text{Track}_\mathrm{cPIT}$.

\subsection{Joint SELD Task}
Several methods for the SELD task are compared with results shown in Table \ref{tab:perf_seld}. No-PS, Hard-PS, and Soft-PS using cross-stitch are the methods shown in Fig. \ref{fig:parameter_sharing}. Among these methods, ACCDOA is an exception which only trains using DoA loss \cite{shimada2020sound}. The proposed EINV2 method is a combination of Soft-PS and the trackwise output format. The second part with \textit{DA} in the table shows the performance with rotation \cite{Mazzon2019} and SpecAugment \cite{park2019specaugment} data-augmentation methods. 

It can be seen in the upper part of Table \ref{tab:perf_seld} that the trackwise output format achieves consistently lower $L E_{\mathrm{CD}}$ and better overall performance than the SELDnet output format for every method. When using the SELDnet output format, No-PS outperforms Hard-PS, which indicates that hard parameter-sharing may be worse than learning two separate tasks, that SED and DoA estimation can be detrimental to each other. It is a bit surprising that ACCDOA using the SELDnet output format outperforms No-PS and Hard-PS using the trackwise output format. This may be because ACCDOA turns the joint task into a single task, which eliminates the detriment caused by joint learning. However, ACCDOA still suffers from a performance loss in SED compared with $\text{SED}_\mathrm{only}$. It may be because ACCDOA uses mean-square-error instead of binary cross entropy as the loss to train SED. It is also challenging for ACCDOA to adapt to cases with different location radii. Among all compared methods, the proposed EINV2 shows the best performance without a compromise compared with learning the subtasks separately. All EINV2 scores are better or equal to separate tasks. This indicates that by using an effective joint learning scheme, SELD can be trained together with mutual benefits.

It can be seen in the lower part of Table \ref{tab:perf_seld} that, when two data augmentation methods are applied, EINV2 outperforms other methods and is comparable with the best ensemble models, which use complex models and more data augmentation methods. EINV2 is a single model with basic VGG-style modules that can be extended to some other networks, such as ResNet or DenseNet.

\section{Conclusion}
\label{sec:conclusion}

We have presented an improved Event Independent Network V2 for sound event localization and detection. It addresses two problems. First, a trackwise output format is able to detect sound events of the same type but with different DoAs. Both frame-level and chunk-level permutation-invariant training are used to solve the track permutation problem. Multi-head self-attention is used to separate predictions in different tracks. Second, a soft parameter-sharing scheme is adopted for joint SELD without a performance compromise compared with learning the subtasks separately. Experimental results show that the proposed EINV2 outperforms previous methods by a large margin. With a single VGG-style model used, EINV2 is comparable with the best ensemble models that use more data augmentation methods.



\section{Acknowledgement}
\label{sec:ack}
This work was supported in part by EPSRC Grants EP/P022529/1, EP/N014111/1 ``Making Sense of Sounds'', EP/T019751/1 ``AI for Sound'', National Natural Science Foundation of China (Grant No. 11804365), and EPSRC grant EP/N509772/1, ``DTP 2016-2017 University of Surrey''.


\bibliographystyle{IEEEbib}
\bibliography{refs}
\end{sloppy}
\end{document}